 \documentclass[final,3p,times,twocolumn]{elsarticle}
\usepackage{graphicx}
\usepackage{braket}
\usepackage{amsmath}
\usepackage{mathtools}
\usepackage{graphics}
\journal{Comptes Rendus Physique}

\begin{document}

\begin{frontmatter}



\title{Quantum trajectories of superconducting qubits}


\author[label1]{S. J. Weber\corref{cor1}}
\cortext[cor1]{Current address: MIT Lincoln Laboratory, 244 Wood Street, Lexington, Massachusetts 02420, USA.}
\address[label1]{Quantum Nanoelectronics Laboratory, Department of Physics, University of California, Berkeley CA 94720, USA.}
\author[label2]{K. W. Murch}
\address[label2]{Department of Physics, Washington University, St. Louis, Missouri 63130, USA. }
\author[label1]{M. E. Schwartz}
\author[label3]{N. Roch}
\address[label3]{ CNRS and Universit\'e Grenoble Alpes, Institut N\'eel, 38042 Grenoble, France.}
\author[label1]{I. Siddiqi}

\begin{abstract}

In this review, we discuss recent experiments that investigate how the quantum sate of a superconducting qubit evolves during measurement.  We provide a pedagogical overview of the measurement process, when the qubit is dispersively coupled to a microwave frequency cavity, and the qubit state is encoded in the phase of a microwave tone that probes the cavity.  A continuous measurement record is used to reconstruct the individual quantum trajectories of the qubit state, and quantum state tomography is performed to verify that the state has been tracked accurately.  Furthermore, we discuss ensembles of trajectories, time-symmetric evolution, two-qubit trajectories, and potential applications in measurement-based quantum error correction.  

\end{abstract}


\end{frontmatter}



 \section{Introduction}
 \label{sec:intro}

The standard description of quantum mechanics considers the time-evolution of isolated quantum systems whose unitary dynamics are governed by the  Schr\"{o}dinger equation.  Measurement is treated as an instantaneous non-unitary process through which a quantum system is projected into an eigenstate of the measured observable with a probability given by Born's rule.  In reality, no system is completely isolated from its environment, and measurements are never truly instantaneous, but occur over some finite timescale determined by the details of the interaction between the measured system and its environment.  The theory of quantum trajectories \cite{carmbook, gardinerbook} considers measurement as a continuous process in time, describing how the state of the quantum system evolves during measurement. 

Due to the intrinsic quantum fluctuations of the environment, measurement is an inherently stochastic process.  If a quantum system starts in a known quantum state $\ket{\psi(0)}$, then by accurately monitoring the fluctuations of its environment it is possible to reconstruct single quantum trajectories $\ket{\psi(t)}$, which describe the evolution of the quantum state in an individual experimental iteration.  

The concept of quantum trajectories was first developed in the early 1990's as a theoretical tool to model continuously monitored quantum emitters \cite{carmbook, dali92,gard92}.  For the next decade, quantum trajectories were used primarily in the quantum optics community, as a theoretical tool for numerical simulations of the ensemble behavior of open quantum systems \cite{dali92,scha95}.  Typically, the master equation of an open quantum system cannot be solved analytically, and thus numerical solutions are often necessary.  For a Hilbert space of dimension $N$, the density matrix $\rho$ consists of $N^2$ real numbers, and the computational time required to solve for its time evolution through the master equation scales as $N^4$ \cite{wisebook}. In contrast, the pure quantum state $\ket{\psi(t)}$ of an individual quantum trajectory can be described by $N$ complex numbers. Therefore, it is often advantageous to simulate an ensemble of stochastic quantum trajectories, which can be averaged together to recover the evolution of the density matrix, $\rho(t)$. Although the formalism of quantum trajectories is constructed from standard quantum mechanics \cite{brun02}, it can provide insight into foundational questions such as the quantum measurement problem \cite{gisi84, dios88, gisi92} and bears a close resemblance to the consistent histories interpretation of quantum mechanics \cite{grif84}.  

Despite widespread theoretical use, quantum trajectories have only been investigated in a handful of experiments, due in part to the difficulty of performing highly efficient continuous quantum measurements. The earliest experiments to continuously monitor individual quantum systems were in the regime of strong measurement, where the system is quickly projected into an eigenstate of measurement, destroying any information about the phase of a coherent superposition.  In such experiments, it is possible to track the `quantum jumps' between eigenstates \cite{ nago86,saut86,berg86, vija11}.  Cavity quantum electrodynamics (CQED) experiments with Rydberg atoms have explored the weak measurement regime, tracking the quantum trajectories of a cavity field as it collapses from a coherent state into a photon number eigenstate \cite{guer07}.  Other CQED experiments have used a cavity probe to continuously track the position of individual Cesium atoms  \cite{hood00}.  Quantum trajectories were first considered for solid state systems in the context of a quantum dot qubit monitored in real time by a quantum point contact charge sensor \cite{koro99, goan01}.  In 2007, the conditional measurement dynamics of a quantum dot were investigated experimentally \cite{sukh07}.  More recently, quantum trajectory theory has been used to solve for the conditional evolution of a continuously monitored superconducting qubit \cite{gamb08,koro11}.  These results, when combined with recent advances in nearly-quantum-limited parametric amplifiers, which can be used to achieve highly efficient qubit readout, have enabled a detailed investigation of measurement backaction \cite{hatr13, camp14}.  

In this article, we review recent experiments \cite{murc13,webe14,roch14,tan14} which, by weakly probing the field of a microwave frequency cavity containing a superconducting qubit, track the individual quantum trajectories of the system.  These are the first experiments, on any system, which use quantum state tomography at discrete times along the trajectory to verify that the qubit state has been faithfully tracked.  From the perspective of quantum information technology, these experiments demonstrate the great extent to which the process of measurement is understood in this system, and may inform future efforts in measurement-based feedback \cite{sayr11, vija12, dela14, groe13} for state stabilization \cite{blok14} and error correction.   

The review is organized as follows.  In section 2, we present a physical picture for continuous quantum measurement of superconducting qubits. In section 3, we demonstrate how to use a measurement result to reconstruct the conditional qubit state after measurement.  In section 4, we explain how to reconstruct and tomographically verify individual quantum trajectories.  Then, in section 5, we examine ensembles of quantum trajectories to gain insight into qubit state dynamics under measurement.  In section 6, we discuss time-symmetric evolution under quantum measurement, and in section 7 we demonstrate quantum trajectories of a two-qubit system.  Finally, in section 8 we explore potential applications in measurement-based feedback control and continuous quantum error correction.

\section{Continuous measurement of superconducting qubits}
\label{cont}

The experiments discussed in this review use artificial atoms formed from superconducting circuits.  We focus in particular on the transmon circuit \cite{koch07} (Fig. 1A) which is composed of the non-linear inductance of a Josephson junction and a parallel shunting capacitance $C_{\Sigma}$. This circuit is characterized by the Josephson energy scale $E_J \equiv \hbar I_0/2e$ and the capacitive energy scale $E_C \equiv e^2/2C_\Sigma$, where $I_0$ is the junction critical current and $e$ is the elementary charge.  A typical transmon circuit, with $E_J/h \sim 20$ GHz and $E_C/h \sim 200$ MHz, has several bound eigenstates (Fig. 1B) with energies $E_m$, where $m$ is a whole number that indexes the states.  The lowest two levels form a qubit subspace, with transition frequency $\omega_{01}/2\pi \equiv (E_1-E_0)/h \sim 5$ GHz, and the difference in frequency between transitions to successively higher levels is given by the anharmonicity $ \alpha \approx E_C$.  Due to the large $E_J/E_C$ ratio the transmon qubit is insensitive to charge noise, which, when combined with low-loss materials \cite{megr12, chan13} and designs that minimize the participation of surface dielectric loss \cite{paik11}, has allowed for planar qubits with coherence times of many tens of microseconds.  

\begin{figure*}
   \begin{center}
\includegraphics[angle = 0, width = .8\textwidth]{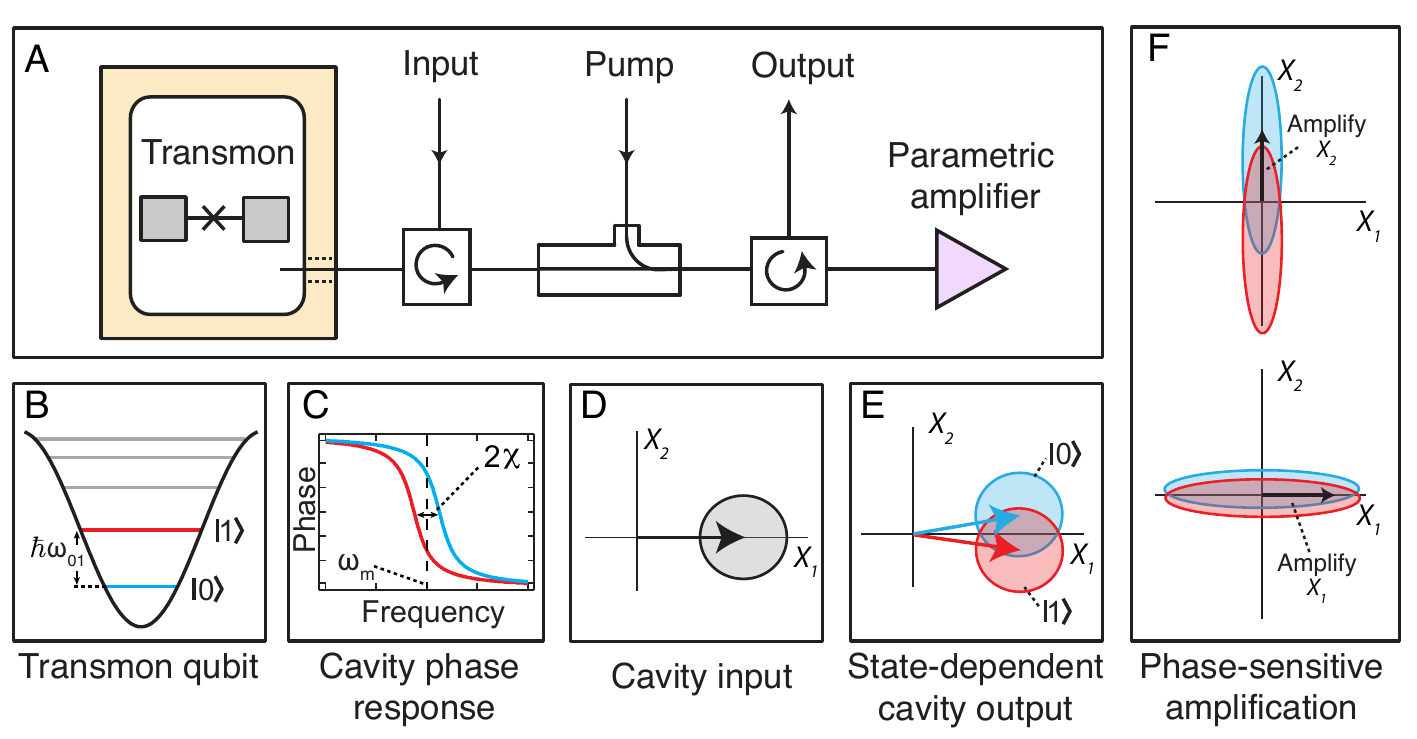}
\end{center}
\caption{\label{fig:fig1} Dispersive measurements of a superconducting qubit.  (A) A transmon qubit couples dispersively to a microwave frequency cavity.  Signals that reflect off of the cavity are amplified by a nearly quantum-limited lumped-element Josephson parametric amplifier.  (B) Schematic representation of the transmon potential and corresponding energy levels.  The two lowest energy levels form the qubit subspace.  (C) Reflected signal as a function of probe frequency.  The resonance frequency is shifted by $2\chi$ depending on whether the qubit is prepared in the $\ket{0}$ or $\ket{1}$ state. (D,E) The cavity is probed with a coherent microwave tone at a frequency $\omega_m$, initially aligned along the $X_1$ quadrature. After leaving the cavity, the tone acquires a quit-state-dependent phase shift.  (F) Phase-sensitive amplification along the $X_2$ (top) and the $X_1$ (bottom) quadratures. }  
\end{figure*}

In order to to control the qubit's interaction with its external environment, it is coupled to the fundamental mode of a three dimensional waveguide cavity of frequency $\omega_c$ at a rate $g$, realizing a cavity quantum electrodynamics (CQED) architecture.  In the dispersive regime, where the qubit-cavity detuning $\omega_q-\omega_r$ is large compared to $g$, the system is described by the Hamiltonian \cite{webe14,blai04} $H = H_0 + H_{\text{int}}+ H_{\text{R}}$, where 

\begin{align}
&H_{\text{int}} = - \hbar \chi a^{\dagger}a \sigma_z \\
&H_{\text{R}} = \hbar \frac{\Omega}{2}\sigma_y.
\end{align}

\noindent Here $H_0$ describes the uncoupled qubit and cavity energies and decay terms, $\hbar$ is the reduced Plank's constant, $\chi$ is the dispersive coupling rate, $a^{\dagger}$ and $a$ are the creation and annihilation operators for the cavity mode, and $\sigma_z$ is the qubit Pauli operator that acts on the qubit state in its energy eigenbasis.  $H_{\text{int}}$ is an interaction term which equivalently describes a qubit-state-dependent frequency shift of the cavity of $-\chi \sigma_z$ (with the $\ket{0}$ state defined as $\sigma_z = +1$) and a qubit frequency that depends on the intracavity photon number $\hat{n} = a^{\dagger}a$ (an a.c. Stark shift).  $H_\text{R}$ describes the effect of an optional microwave drive at the qubit frequency which causes the qubit state to rotate about the $y$ axis of the Bloch sphere at the Rabi frequency $\Omega$.  

Because the $H_{\text{int}}$ commutes with $\sigma_z$, the qubit-state-dependent phase shift can be used to perform continuous quantum non-demolition (QND) measurment of the qubit state in its energy eigenbasis \cite{blai04,bragbook}.  Figure 1C illustrates the phase of the reflected signal as a function of frequency.  If we choose to measure at a frequency $\omega_m = (\omega_{\ket{0}} + \omega_{\ket{1}})/2$, where $\omega_{\ket{0}}$ and $\omega_{\ket{1}}$ are the cavity frequencies when the qubit is in the ground and excited states, respectively, then the phase difference in the internal cavity field for the two qubit states is given by $\Delta \theta = 4|\chi|/\kappa$, where $\kappa$ is the cavity decay rate.  In the experiments presented here, we work in the small phase shift limit, with $|\chi|/\kappa \sim 0.05$.

We probe the cavity by applying a measurement tone at frequency $\omega_m$ initially aligned along the $X_1$ quadrature (Fig. 1D).  Due to the vacuum fluctuations of the electromagnetic field, the quadrature amplitudes $X_1$ and $X_2$ of this field will fluctuate in time.  The circle in Figure 1D represents the Gaussian variance of the input signal time-averaged for a time $\Delta t$.  The area of the circle is inversely proportional to $\Delta t$.  After reflecting off of the cavity, the measurement signal acquires a qubit-state-dependent phase shift, as depicted in Figure 1E. In the small $|\chi|/\kappa$ limit, the $X_2$ quadrature of the reflected signal signal contains information about the cavity phase, which is proportional to the qubit state.  Likewise, the $X_1$ quadrature contains information about the amplitude of the cavity field and thus the fluctuating intracavity photon number.  

In order to track the qubit state through an individual measurement, we need to accurately monitor the quantum fluctuations of the measurement signal, which are typically much smaller than the thermal fluctuations of the room temperature electronics that are needed to record the measurement result.   Therefore, we must first amplify the signal above this noise floor.  State-of-the-art commercial low-noise amplifiers, which are based on high electron mobility transistors (HEMTs) and can be operated at $4$ K, add tens of photons of noise to the measurement signal.  Therefore, a more sensitive pre-amplifier is needed in order to overcome the added noise of the HEMT amplifier.  

Over the past few years, Josephson junction based superconducting parametric amplifiers have emerged as an effective tool for realizing nearly-quantum-limited amplification.  Phase-preserving amplifiers such as the Josephson parametric converter \cite{berg10} amplify both quadrature amplitudes evenly by a factor of $\sqrt{G}$, where $G$ is the power gain of the phase-preserving amplifier, and add at least a half photon of noise \cite{cave82} to the signal.  Here, we focus instead on phase-sensitive amplification from a lumped-element Josephson parametric amplifier \cite{hatr11}, where one quadrature is amplified by a factor of $2\sqrt{G}$ and the other quadrature is de-amplified by the same factor.  

When we apply a coherent measurement tone, characterized by an average intracavity photon number $\bar{n}$, its quantum fluctuations will cause the phase coherence of a qubit superposition state to decay at the ensemble dephasing rate $\Gamma = 8\chi^2\bar{n}/\kappa$.  The ensemble dephasing rate will be the same regardless of how we choose to process the measurement signal after it leaves the cavity.  However, the backaction of an $\emph{individual}$ measurement will depend significantly on our choice of amplification scheme. 

As depicted in Figure 1F, after a measurement tone initially aligned along the $X_1$ quadrature reflects off of the cavity and acquires a qubit-state-dependent phase shift and is then displaced to the origin of the $X_1\text{-}X_2$ plane by a coherent tone, we can choose to either amplify the $X_2$ quadrature (top panel) which contains qubit-state information or the $X_1$ quadrature (bottom panel) which contains information about the fluctuating intracavity photon number. Consider a qubit initially prepared in an equal superposition of $\sigma_z$ eigenstates, say $\sigma_x = +1$.  If we perform ideal phase-preserving amplification of $X_1$, we also de-amplify the photon number information, and the measurement backaction drives the qubit state along a meridian of the Bloch sphere toward one of the poles. We refer to this case as a  $z-$measurement, because we acquire information about the qubit state in the $\sigma_z$ basis.  If instead we amplify $X_2$, we also de-amplify the qubit-state information, and measurement backaction drives the qubit state along the equator of the Bloch sphere.  We refer to this case as a $\phi-$measurement, because we can track the phase of a qubit superposition state over the course of an individual measurement.  For phase-preserving amplification both types of backaction are present.  

We first focus on the case of the $z-$measurement.  After the measurement tone leaves the parametric amplifier and passes through further stages of amplification we demodulate the signal and record the $X_2$ quadrature amplitude as a digitizer voltage $V(t)$.  For a measurement of duration $\Delta t$, the measurement outcome $V_m$ is given by the time-average of $V(t)$.  Depending on whether the qubit is initially prepared in the ground of the excited state, the Gaussian distribution describing the probability attaining a particular measurement outcome will be shifted by a voltage $\Delta V \propto \Delta \theta$.  In this review, we define a dimensionless measurement outcome  $r = 2 V_m / \Delta V$, such that the ground and excited state distributions are centered about $r = \pm 1$, respectively, as illustrated in Figure 2A,D. From these distributions, we define the dimensionless measurement strength $S \equiv (2/a)^2$, where $a$ is the standard deviation of the dimensionless measurement distributions, which scales as $(\Delta t)^{-1/2}$.  We also define the characteristic timescale $\tau$ over which the qubit state is projected as the amount of time required for the measurement histograms to be separated by twice their standard deviation, $\tau \equiv 4 \Delta t / S$.  When $S$ is large ($\Delta t \gg \tau$), the ground and excited state histograms are well separated (Fig. 2A), and it is possible to determine the qubit state with high fidelity in an individual measurement.  The measurement projects the qubit into an energy eigenstate, where it will remain after measurement.  Instead, if $S$ is small ($\Delta t \lesssim \tau$), the ground and excited state histograms overlap (Fig. 2D), and an individual measurement only partially projects the qubit state.  

\begin{figure*}
   \begin{center}
\includegraphics[angle = 0, width = 1\textwidth]{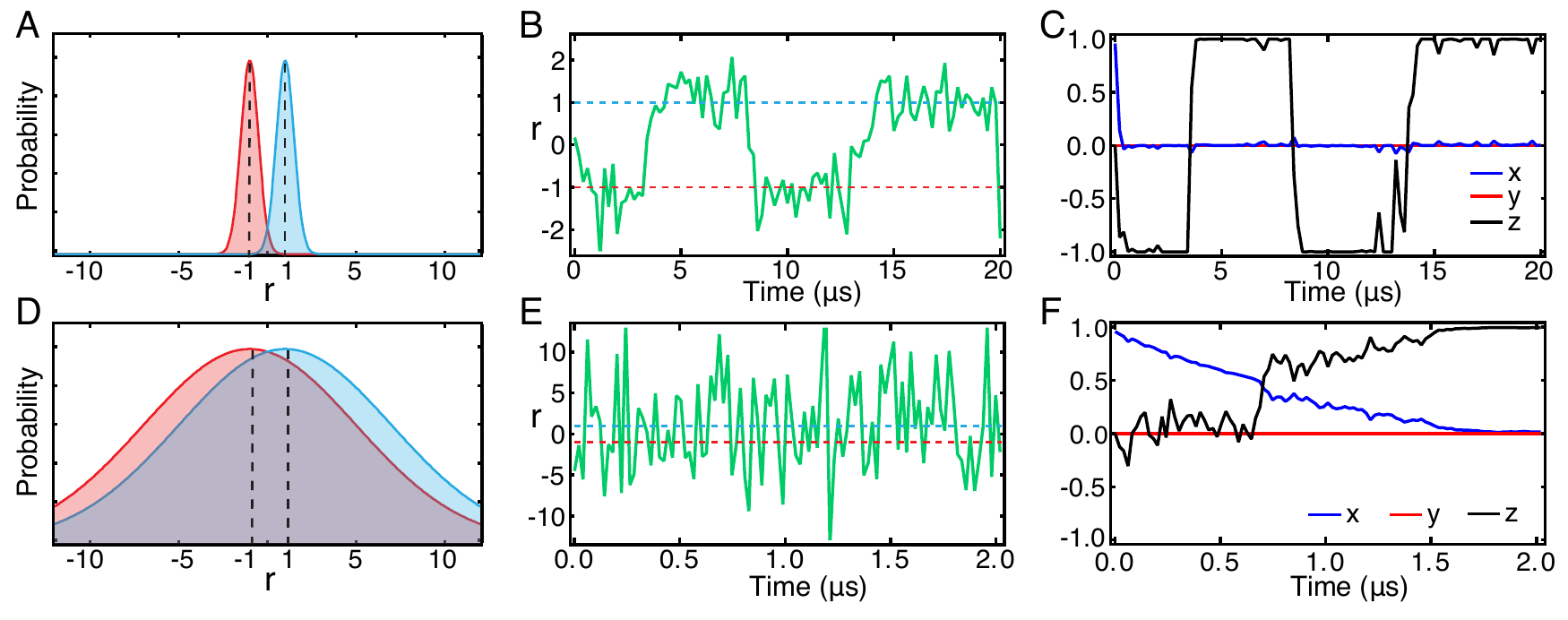}
\end{center}
\caption{\label{fig:fig2} Continuous quantum measurement. Illustrative measurement histograms for a single time-step (A,D) along with simulated measurement records (B,E) and corresponding quantum trajectories (C,F).  In the top panels, $\Delta t = 200$ ns, $\tau = 50$ ns, and $\Omega/2\pi = 8$ MHz, illustrating the quantum jumps regime.  In the bottom panels, $\Delta t = 20$ ns, $\tau = 150$ ns, and $\Omega = 0$, illustrating the diffusive regime.}  
\end{figure*}

Formally, general quantum measurements (partial or projective) are described by the theory of positive operator-valued measures (POVM) which yield the probability $P(m)=\mathrm{Tr}(\Omega_m \,\rho\, \Omega_m^\dagger)$ for obtaining an outcome $m$, and the associated back action on the quantum state, $\rho \rightarrow \Omega_m \,\rho\, \Omega_m^\dagger/P(m)$, where the operators $\Omega_m$ obey  $\sum_m \Omega^\dagger_m \Omega_m = \hat{I}$.  For example, for the projective measurements $\Omega_{\pm z} = (\hat{I} \pm \sigma_z)/2$, the probability of a measurement yielding the qubit in the $+z$ state is $P(+z) = \mathrm{Tr}(\Omega_{+z} \rho \Omega_{+z})= (1+\langle \sigma_z\rangle)/2$. The partial measurements discussed in this review are described by the POVM \cite{wisebook, jaco06},

\begin{align}
%\Omega_V = \left(8 k \eta dt/\pi \right)^{1/4} e^{(-4 \eta k dt (V- \sigma_z/2)^2)}
\Omega_r = \left(2 \pi a^2 \right)^{-1/4} e^{(-(r- \sigma_z)^2/4a^2)}
\label{povm}
\end{align}

\noindent where,  $1/4a^2 =  \Delta t/ 4 \tau$. The $\sigma_z$ term in $\Omega_r$ causes the back action on the qubit degree of freedom, $\rho \rightarrow \Omega_r \rho \Omega_r^\dagger$, due to the readout of the measurement result $r$, resulting in the measurement dynamics discussed in this review.

Our ability to reconstruct the qubit state after an individual partial measurement is determined by the measurement quantum efficiency $\eta_m$.  We have established that the noisy measurement tone contains information about the qubit state and cause an ensemble dephasing rate of $\Gamma$.  In general, only a fraction, $\eta_m$ of this information is experimentally accessible, and the remainder is lost to environmental degrees of freedom.  The measurement efficiency can be reduced from its ideal value of $\eta_m =1$ by losses between the cavity and the parametric amplifier, described by the collection efficiency $\eta_{col}$ and by added noise in the amplification chain, described by an the amplification efficiency $\eta_{amp}$.  In the experiments presented here, $\eta_m = \eta_{col} \eta_{amp} \sim 0.4$.  The measurement strength depends linearly on $\eta_m$, and for dispersive measurements in the small phase shift limit is given by $S = 64 \chi^2 \bar{n} \eta_m \Delta t/ \kappa$.

We now turn our attention to continuous quantum measurement.  In Figure 2 we illustrate quantum trajectories in the limiting cases of strong and weak measurement.  We consider a sequence of $n$ measurements occurring at times $\{t_k = k \Delta t\}$ for $k = 0,1,...,n-1$, which result in a set of dimensionless measurement results $\{r_k\} = \{r_0,r_1,...,r_{n-1}\}$. If the qubit is prepared in a known initial state, then we can use the measurement results to track the qubit state as it evolves under measurement, computing the set of conditional qubit states $\{q_k\} = \{q_0,q_1,...,q_{n-1}\}$ corresponding to an individual measurement record $\{r_k\}$.  Here the Bloch vector $q = (x,y,z)$ describes a general mixed single-qubit state in terms of the the components $x \equiv \text{tr}[\rho\hat{\sigma}_x]$, $y \equiv \text{tr}[\rho\hat{\sigma}_y]$, and $z \equiv \text{tr}[\rho\hat{\sigma}_z]$, where $\rho$ is the qubit density matrix.   In the limit where $\tau \lesssim \Delta t$, each time-step constitutes a (nearly) projective measurement.  In the absence of any non-measurement dynamics, after the first time-step subsequent measurements will continue to project the qubit into the eigenstate corresponding to the initial measurement result. However, any additional dynamics, such as energy relaxation or Rabi driving, which occur on a timescale faster than $\Delta t$ will result in discontinuous jumps in the measurement record corresponding to quantum jumps \cite{vija11} of the qubit state (Fig. 2B,C).  

In the opposite limit, where $\Delta t \ll \tau$, each measurement will only slightly perturb the qubit state.  However, by performing a sequence of repeated partial measurements such that $t_{n-1} \gg \tau$, we can realize a projective measurement.  In this case, the noisy detector signal can be used to reconstruct the diffusive trajectory of the qubit state as it is gradually projected toward a measurement eigenstate (Fig. 2E,F).  

\section{Reconstructing the conditional quantum state}
\label{cond}

In this section, we discuss in detail how to reconstruct the qubit state conditioned on an individual measurement outcome.  One approach, taken in references \cite{roch14, tan14} is to solve a stochastic master equation for conditional qubit state.  Here, we instead describe phenomenological approach based on a Bayesian statistics \cite{koro11}, which provides a particularly simply approach to single-qubit trajectories and was used in references \cite{murc13} and \cite{webe14}.  As recently demonstrated in reference \cite{tan14}, for the a single qubit under weak measurement and weak Rabi driving, both approaches yield similar results.

Consider a qubit prepared in the initial state $\rho(t = 0)$, which is weakly measured for a time $\Delta t$, yielding a measurement result $r$.  Here, we show how to apply Bayes rule of conditional probabilities to update our knowledge of the qubit state after the measurement.  We first focus on the case of a $z-$measurement, with $\Omega = 0$.  From Bayes rule, we have

\begin{align}
\label{prob}
P(i|r) =&\, \dfrac{P(r|i)P(i)}{P(r)},
\end{align}

\noindent where $i$ describes the basis states $\{\ket{0},\ket{1}\}$.  Here, the initial probabilities for finding the qubit in the ground or excited states are given by $P(0) = \rho_{00}(t\! =\!0)$ and $ P(1) = \rho_{11}(t\! = \!0)$. By expressing the measurement distributions $P(r|i)$ explicitly and taking the ratio of the conditional probabilities in equation \eqref{prob} for both basis states, we find that 

\begin{eqnarray}
\label{diag}
\frac{\rho_{11}(\Delta t)}{\rho_{00}(\Delta t)} = \frac{\rho_{11}(0)}{\rho_{00}(0)}\frac{\text{exp}[-(r+1)^2/2a^2]}{\text{exp}[-(r-1)^2/2a^2]}.
\end{eqnarray}

\noindent For a qubit initially prepared in the state $q_I = (1,0,0)$ we find that

\begin{eqnarray}
\label{zz}
z^z= \text{tanh}\left(\frac{r\Delta t}{\tau}\right).  
\end{eqnarray}  

\noindent where the superscript `$z$' denotes a $z-$measurement.

Note that thus far we have used a classical rule of conditional probabilities to determine how the qubit populations evolve under measurement.  Following reference \cite{koro11}, we account for the qubit coherence through the phenomenological assumption that 

\begin{eqnarray}
\label{xz}
x^z = \sqrt{1-(z^z)^2}e^{-\gamma \Delta t}
\end{eqnarray}

\noindent Here the first term enforces normalization and the second term reflects our imperfect knowledge of the environment and leads to qubit dephasing characterized by the rate $\gamma = \Gamma - 1/2\tau$, where $\Gamma = \Gamma + 1/T_2^*$ is the ensemble dephasing rate and $T_2^*\sim 20 \,\mu$s  is the characteristic timescale for extra environmental dephasing. 

For the case of a $\phi-$measurement, $z$ remains zero, and $x$ and $y$ are periodic in the accumulated qubit phase shift, and are given by \cite{koro11}

\begin{align}
\label{xp}&x^{\phi} = \text{cos}\left(\frac{r \Delta t}{\tau}\right)e^{-\gamma \Delta t}, \\
\label{yp}&y^{\phi} = -\text{sin}\left(\frac{r \Delta t}{\tau}\right)e^{-\gamma \Delta t}, \\
\end{align}

\noindent where the superscript `$\phi$' denotes a $\phi-$measurement.  Figure 3 illustrates the conditional quantum state as a function $r$ for a $z-$measurement (panel A) and a $\phi-$measurement (panel B), with $\tau = 600$ ns and $\Delta t = 400$ ns.  Note that the dephasing rate $\gamma$ due the unaccessible part of the measurement signal is the same regardless of our choice of amplification axis. In both cases, a measurement outcome of $r = 0$ will leave $y$ and $z$ unchanged, but $x$ is reduced by a factor of $\text{Exp}[-\gamma \Delta t]$.

A useful feature of dispersive CQED measurements is the ability to rapidly tune the measurement strength by changing the amplitude of the measurement tone.  Therefore, it is straightforward to implement experimental sequences which combine partial and projective measurement.  The sequence shown in Figure 3c is used to implement conditional quantum state tomography to verify that we can accurately account for the backaction of an individual measurement.  The qubit is prepared in the initial state $(1,0,0)$, and weakly measured for a time $\Delta t$.  Then, we perform an optional qubit rotation (of $\pi/2$ about the $\hat{y}$ axis to reconstruct $x$, $\pi/2$ about $-\hat{x}$ to reconstruct $y$, and no pulse to reconstruct $z$) followed by a projective measurement. For a given measurement outcome $r$, we perform a tomographic state reconstruction on the sub-ensemble of experimental iterations with similar measurement outcomes, in the range $r\pm\epsilon$, where $\epsilon \ll 1$.  For superconducting qubits, this technique was first introduced in reference \cite{hatr13}, which considers the case of phase-preserving amplification. Shortly thereafter, this technique was demonstrated for phase-sensitive amplification \cite{murc13}.  

\begin{figure}
\begin{center}
\includegraphics[angle = 0, width = .5\textwidth]{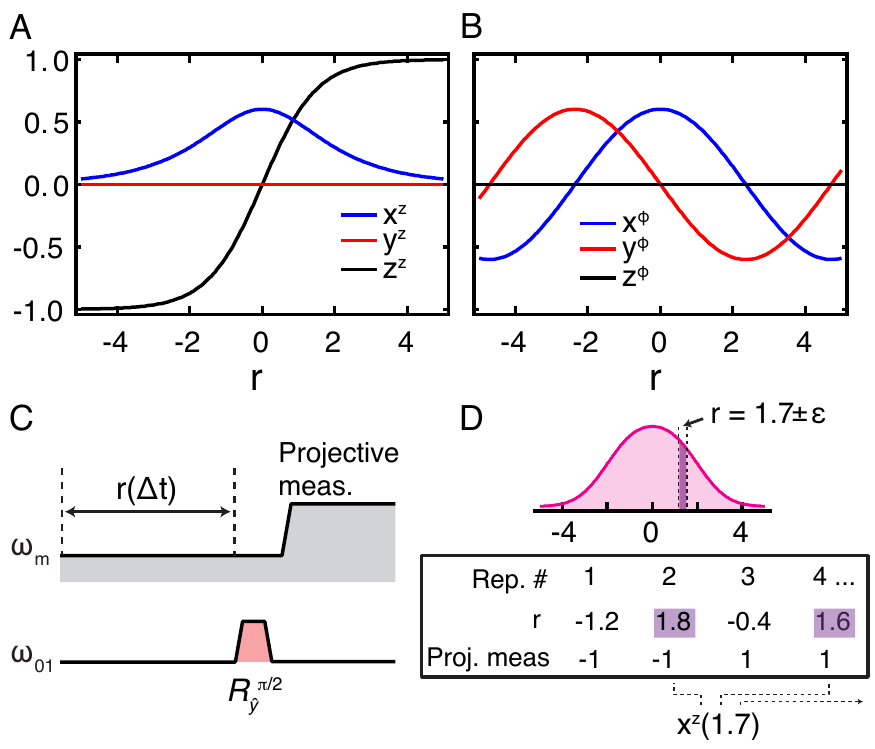}
\end{center}
\caption{\label{fig:fig2} Reconstructing the conditional quantum state.  (A,B) The conditional quantum state after a measurement result $r$ for a qubit initially prepared in the state $(1,0,0)$, with $\Delta t = 400$ ns, $\tau = 600$ ns, and $\gamma = 1.3*10^6 \text{s}^{-1}$.  Panel A depicts a $z-$measurement, and panel B depicts a $\phi$ measurement.  (C) Experimental sequence for reconstructing the $x$ component of the conditional quantum state.  (D) To perform quantum state tomography conditioned on the measurement result $r=1.7$, we average together the projective measurement outcomes for the sub-ensemble of measurement outcomes where  $r = 1.7 \pm \epsilon$.}  
\end{figure}

\section{Tracking individual quantum trajectories}
\label{traj}

Consider a qubit initially prepared in a known state $q_I$, which undergoes a sequence of $n$ partial measurements with outcomes $\{r_k\}$, as described in section \ref{cont}.  In the limit where the duration $\Delta t$ of each measurement approaches zero, the set of conditional states $\{q_k\}$ describes a the quantum trajectory $q(t)$.  For simplicity, from here on we restrict our discussion to the case of a $z-$measurement.  When $\omega = 0$, we can calculate the conditional quantum state at each time-step $t_k = k \Delta t$ from equations \eqref{zz} and \eqref{xz} using only the initial state and the time-averaged measurement signal $\bar{r} = 1/k \sum_{k=0}^{k-1}r_k$.  However, when $\Omega > 0$ the measurement dynamics do not commute with the Rabi drive, and therefore the order of the measurement outcomes matters, and $\bar{r}$ no longer contains sufficient information to reconstruct the quantum trajectory.  Instead, if $\Delta t \ll \Omega$ we can perform a sequential two-step state update procedure introduced in reference \cite{webe14}.  For each time-step $t_k$, we calculate $q_k$ by first applying a Bayesian update to the state $q_{k-1}$ to account for the measurement result $r_k$, and then by applying a unitary rotation to account for the the Rabi drive during the time $\Delta t$.  Example quantum trajectories are shown in Figure 4A,B for $\Omega/2\pi = 0$ and $0.4$ MHz, respectively, and $\tau = 1.28 \, \mu$s.  The corresponding ensemble average evolution is shown in panels C and D.

\begin{figure}
\begin{center}
\includegraphics[angle = 0, width = .5\textwidth]{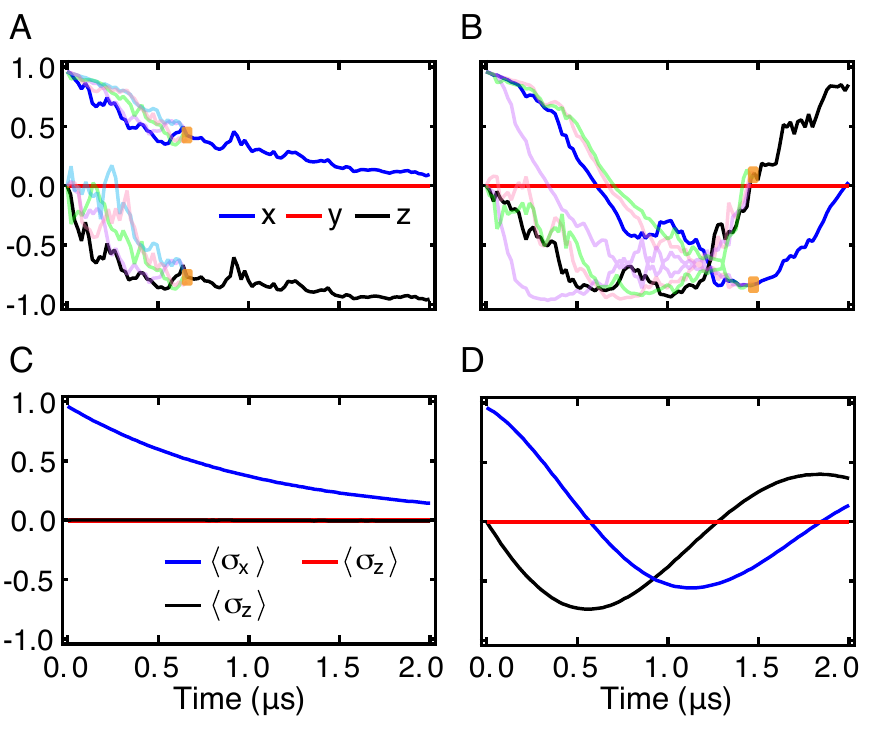}
\end{center}
\caption{\label{fig:fig4} Reconstructing individual quantum trajectories.  Here, $\tau = 1.28 \, \mu$s, $\gamma =  2.7\times 10^{-7}s^{-1}$, and  $\Omega/2\pi = 0$ (A, C) and $0.4$ MHz (B,D).  Panels A and B depict the ensemble average evolution.  Panels C and D display simulated individual quantum trajectories ending at $t_n = 2 \, \mu$s, with the $x, y$, and $z$ components depicted in blue, red, and black, respectively. The orange regions represent a matching window of $\epsilon = 0.05$ at $t = 0.66 \, \mu$s (C) and $1.48 \, \mu$s (D). Sample trajectories that end within the matching window are shown in other colors.  }  
\end{figure}

While pervious experiments in other systems have reconstructed individual diffusive quantum trajectories \cite{guer07}, reference \cite{murc13} was the first to use conditional quantum state tomography to verify that the trajectories were reconstructed accurately.  Here, we present a brief outline of the tomographic validation procedure.  We perform a large number of experimental iterations ending at different times $t_f$, which are followed by a qubit rotation and a projective measurement. We use a single full-length experimental iteration (with $t_f = (n-1) \Delta t$ to generate a target trajectory, denoted $\tilde{q}(t) \equiv (\tilde{x}(t),\tilde{y}(t), \tilde{z}(t))$.  Then, for each experimental sequence of total measurement duration $t_k$ (and a given orientation of tomography pulse), we compute the quantum trajectory $q(t)$.  We perform conditional quantum state tomography separately at each time $t_k$ using the subset of experimental iterations with $x(t_k) = \tilde{x}(t_k)\pm\epsilon$ and $z(t_k) = \tilde{z}(t_k)\pm\epsilon$ where $\epsilon \ll 1$, and we have assumed that $y=0$.  The orange shaded regions in panels C and D of Figure 4 represent matching windows at $t_k = 0.66 \, \mu$s and $1.28 \, \mu$s, respectively.  Trajectories which fall within the matching window at $t_k$ are used in the tomographic reconstruction of $q(t_k)$.   

\section{Distributions of trajectories}
\label{dist}

By tomographically reconstructing individual quantum trajectories, as discussed above and initially demonstrated in reference \cite{murc13}, we have proven that we can accurately track the qubit state over the course of any individual measurement.  In this section, we consider how quantum trajectory experiments are useful for building an intuition for how the qubit state is most likely to evolve under measurement. As discussed in reference \cite{webe14}, distributions of quantum trajectories offer a convenient qualitative tool for visualizing the interplay between measurement dynamics and unitary evolution.  

The greyscale histograms in panels A and B of Figure 5 display the simulated distribution of quantum trajectories for $\tau = 1.28 \, \mu$s, $\gamma =  2.7\times 10^{-7}s^{-1}$, and  $\Omega/2\pi = 0.4$ MHz.  Note that due to the Rabi drive, the measurement initially projects the qubit preferentially toward the excited state ($z = -1$).  At intermediate times a wide range of qubit states are possible, and after half a Rabi period the qubit is preferentially projected toward the ground state.  In experiments with superconducting qubits, $\tau$ and $\Omega$ can be readily tuned, and distributions of quantum trajectories are experimentally accessible for a wide range of parameters.  

It is also possible to consider the conditional quantum dynamics of the sub-ensemble of trajectories which end in a particular quantum state, or post-selection.  Panels C and D of Figure 5 display the distribution of trajectories which end in the final state $x_F = 0.1 \pm .08$, $z_F = 0.55 \pm .08$.  By analyzing the statistical properties of such distributions, it is possible to answer questions of broad interest in the field of quantum control.  The experiments of reference \cite{webe14} focus on one such question: what is the most probable path through quantum state space connecting an initial state  $\ket{\psi_i}$ and a final state $\ket{\psi_f}$ in a given time $T$?

One straightforward theoretical approach to this problem would be to solve the stochastic master equation (SME) numerically for a large ensemble of repetitions, and then to perform statistical analysis on the sub-ensemble of trajectories which end in $\ket{\psi_f}$ at time $T$.  An alternative approach based on an action principle for continuous quantum measurement was developed in reference \cite{chan13b}. The action principle naturally incorporates post-selection and yields a set of ordinary (non-stochastic) differential equations for the most probable path which are simper to solve numerically than the full SME.  In the limit of no post-selection, this approach is consistent with SME formulation. The results of reference \cite{webe14} experimentally verify the predictions of this action principle for the case of a single qubit under simultaneous measurement and Rabi drive.  However, the action principle is a general theory which can be applied to a wide variety of quantum systems and may prove useful in designing optimal quantum control protocols.  

\begin{figure}
\begin{center}
\includegraphics[angle = 0, width = .48\textwidth]{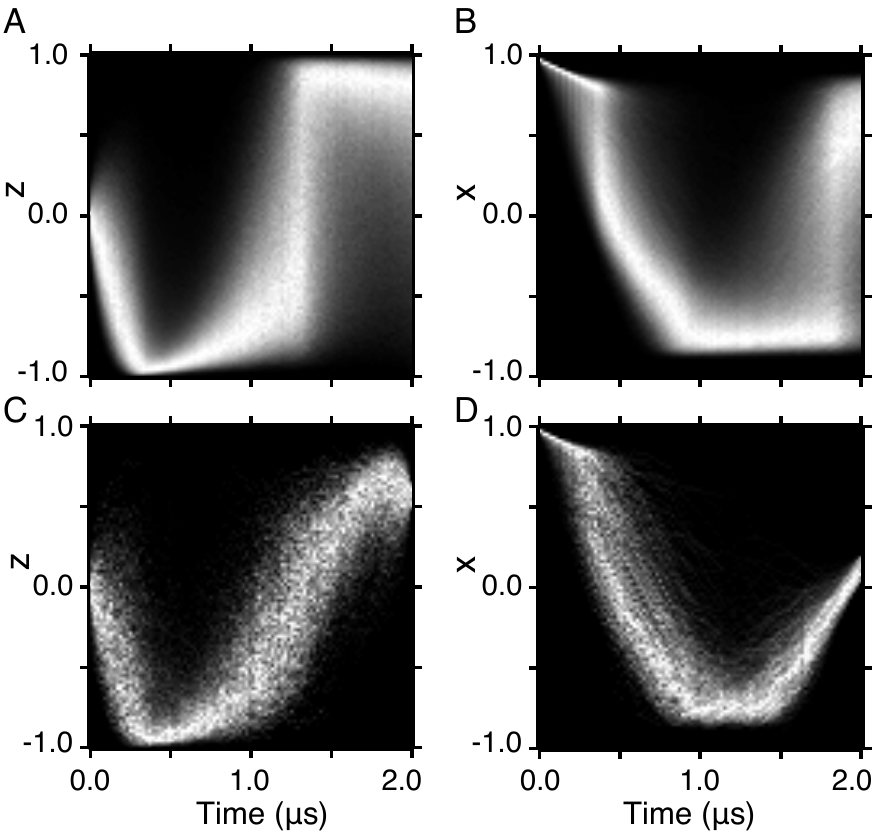}
\end{center}
\caption{\label{fig:fig4} Greyscale histograms of quantum trajectories based on $5\times 10^{4}$ simulated trajectories.  Here, $\tau = 1.28 \, \mu$s, $\gamma =  2.7\times 10^{-7}s^{-1}$, and  $\Omega/2\pi = 0.4$ MHz.  Histograms are normalized such that the most frequent value at each time point is $1$.  Panels A and B depict the full distribution of $x$ (A) and $y$ (B) trajectories.  Panels C and D display the sub-ensemble of trajectories which end in the final state $x_F = 0.1 \pm .08$, $z_F = .55 \pm .08$.}  
\end{figure}

\section{Time-symmetric state estimation}
\label{time}

We have so far focused on the use of the quantum state as a predictive tool; the quantum trajectories presented in the previous sections describe the evolution of expected average outcome of observables $\sigma_x,\sigma_y,\sigma_z$,  which relate in a straightforward way to the probability of obtaining a certain outcome in subsequent projective tomography measurements.  However, it is also possible to follow the qubit state evolution \emph{backward} in time to predict an unknown measurement result from the past. 
 
Consider the following guessing game: Two experimenters can perform measurements on the same quantum system. At a time $t$ the first experimenter makes a measurement of some observable $\Omega_m$ and hides the result.  The second experimenter then must guess the outcome $m$ that the first experimenter received.  If the second experimenter only has access to the quantum system's state before the first experimenter's measurement, then the theory of POVMs provides the second experimenter with probability for each outcome $m$ and the ability to make the best possible guess.  However, if the second experimenter is allowed to probe the quantum system after the first experimenter has conducted her measurements, can he make a better prediction for the hidden result?  Indeed, since more information about the system is available at a later time the second experimenter can make more confident predictions.   It can be shown \cite{wise02, gamm13} that the probability of the outcome $m$ is given by,
\begin{eqnarray}
P_p(m)  = \frac{\mathrm{Tr}(\Omega_m \rho_t \Omega_m^\dagger E_t)}{\sum_m \mathrm{Tr}(\Omega_m \rho_t \Omega_m^\dagger E_t)}, \label{eq:pqs}
\end{eqnarray}
where $\rho_t$ is the system density matrix at time $t$, conditioned on previous measurement outcomes and propagated forward in time until time $t$, while $E_t$ is a matrix which is propagated backwards in time in a similar manner and accounts for the time evolution and measurements obtained after time $t$. The subscript $p$ denotes ``past'', and in \cite{gamm13} it was proposed that, if $t$ is in the past, the pair of matrices $(\rho_t,E_t)$, rather than only $\rho_t$, is the appropriate object to associate with the state of a quantum system at time $t$.  It is worth noting that information from before the first experiment's measurement, which is encoded in $\rho_t$, and information after this measurement (encoded in $E_t$) play a formally equivalent role in the prediction for the experimenter's result. It is natural that full measurement records would contain more information about the system and several precision probing theories \cite{mank09,mank09pra,mank11,arme09,whea10} have incorporated full measurement records.
 
Recent experiments have applied Eq.(\ref{eq:pqs}) to systems with Rydberg atoms \cite{ryba14} and superconducting qubits \cite{tan14}, confirming how full measurement records allow more confident predictions for measurements performed in the past.  This applies to both projective and weak (weak value) measurements and in the case of weak measurements the experiments reveal how the orthogonality of initial ($\rho_t$) and final ($E_t$) states  leads to the occurrence of anomalous weak values \cite{tan14}.

\section{Two-qubit trajectories}
\label{two}

\begin{figure*}
\begin{center}
\includegraphics[angle = 0, width = .78\textwidth]{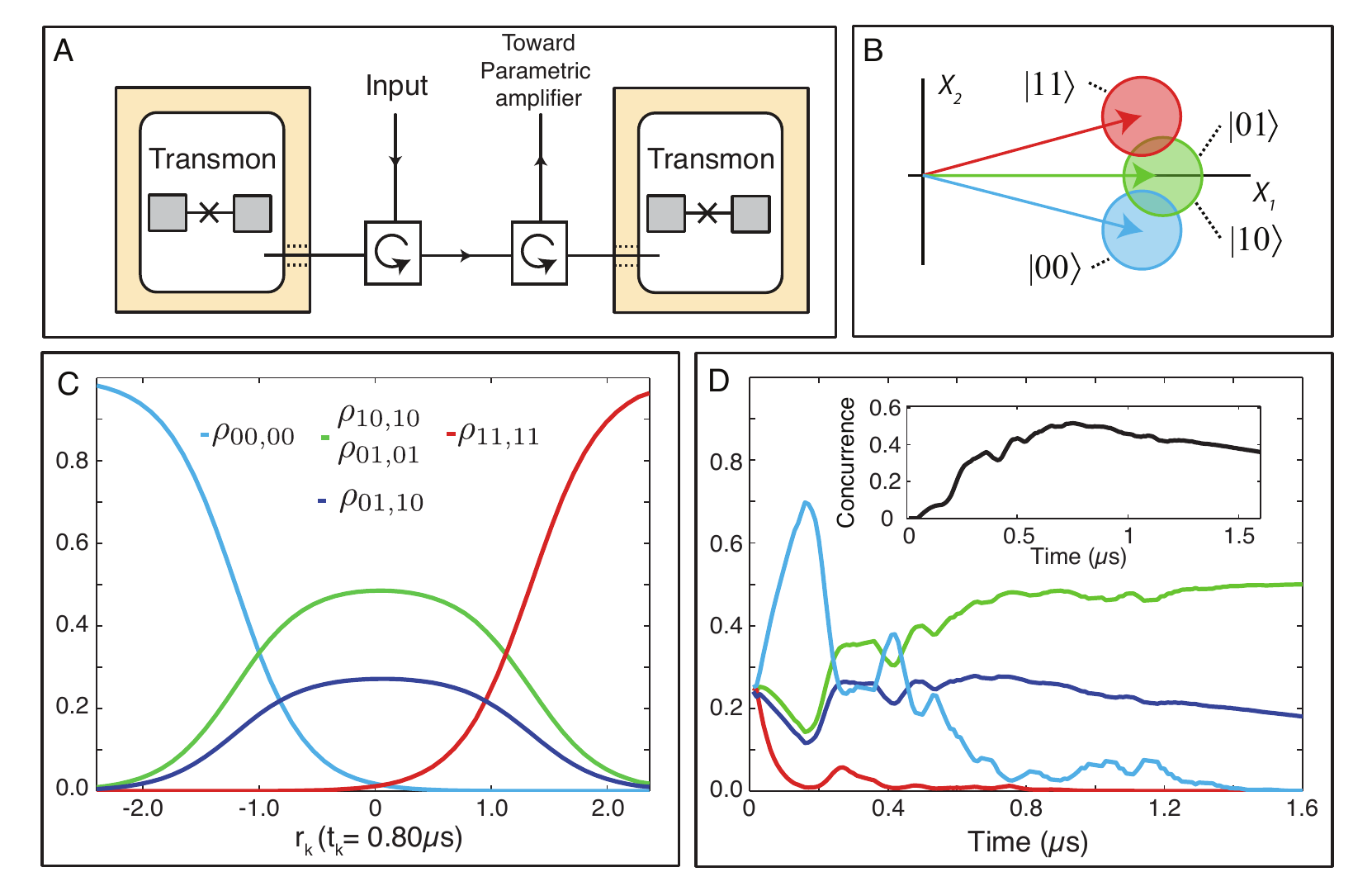}
\end{center}
\caption{\label{fig:fig6} Two-qubit trajectories  (A) Schematic of the measurement setup. (B) The cascaded cavities are probed with a coherent microwave tone, initially aligned along the $X_1$ quadrature. After probing both cavities, the tone can acquire three different phase shifts:  $+\Delta\theta$, $-\Delta\theta$ and $0$ corresponding to the states $\ket{11}$,  $\ket{00}$ and $\ket{01}/\ket{10}$ respectively. (C) Conditional quantum state after a measurement result $r$ of length $t_k=k\Delta t=0.8\mu s$, for both quits initially prepared in the maximally superposed state ($(\ket{0}+\ket{1})/\sqrt(2)\otimes(\ket{0}+\ket{1})/\sqrt(2)$ and $\tau=0.75\mu s$. (D) Single quantum trajectory of the cascaded two-qubit system. The color code is similar to the one used in panel C. The inset shows the corresponding concurrence.}  
\end{figure*}

To this point, we have focused on the trajectories of a single-qubit system.  However, the trajectory formalism is readily extensible for studying the dynamics of a multiple-qubit system.  Such systems are of interest because they allow us to directly observe and study the generation of entanglement on a single-shot basis \cite{will08b}.  In particular, we study a cascaded CQED system comprised of two superconducting qubits housed in two separate cavities that sequentially probed in reflection by a single coherent state (Figure \ref{fig:fig6}). The quantum trajectory formalism for such a system was first developed in 1993 \cite{carmbook}, and was observed experimentally in 2014 \cite{roch14}.  

The Hamiltonian for this cascaded quantum system is given by
\begin{equation}
H = H_0 + \chi_1 a^\dagger a \sigma_z^1 + \chi_2 b^\dagger b \sigma_z^2 - i \frac{\sqrt{\kappa_1\kappa_2 \eta_{loss}}}{2} (a^\dagger b - b^\dagger a),
\end{equation}
where $H_0$ comprises the uncoupled Hamiltonians for the two cavities and qubits; $\chi_i$ are the dispersive shifts; $a$ $(b)$ and $a^\dagger$ $(b^\dagger)$ are the creation and annihilation operators for the first (second) cavity; $\kappa_i$ are the decay rates of the cavities; and $\eta_{loss} \approx 0.8$ represents the transmission efficiency between the two cavities. 

In the multiple-qubit regime, it is more convenient to work in the measurement basis ($\ket{00}$, $\ket{01}$, $\ket{10}$ and $\ket{11}$) rather than the Pauli basis set ($\sigma_i \otimes \sigma_j$), since the measurement in the multi-qubit case does not project along a single-qubit Pauli operator. In the experiments described in reference \cite{roch14}, the sequential measurement realizes a half-parity operation (Fig 5. A,B): the measurement tone acquires distinct phase shifts of $\pm \Delta\theta$ for the even-parity states $\ket{00}$ and $\ket{11}$, and an identical (null) phase shift in the odd parity subspace ($\ket{01}$ and $\ket{10}$).

In cascaded quantum systems, the effect of the losses between the systems is of primary importance and need to be fully taken into account when a quantitative description is needed \cite{roch14}. However in this review, we make the choice to set $\eta_{loss}=1$ (zero losses) for the sake of simplicity. In addition, we make the assumption that the dispersive shifts are equals for the two cavities ($\chi_1=\chi_2=\chi$) as well as the decay rates ($\kappa_1=\kappa_2=\kappa$).

Similarly to the single-qubit case, we can define the measurement outcome $V_k$ as the time-average of the $X_2$ quadrature voltage: $V_k=1/(k\Delta t)\int_0^{k\Delta t}V_{X_2}(t) dt$. The dimensionless measurement outcome is thus given by $r_k=2V_k/\Delta V$ where $\Delta V \propto \Delta\theta$ is defined as the distance between the measured Gaussian histogram centers for $\ket{00}$ and $\ket{01}/\ket{10}$. The measurement realises a projection on a timescale $\tau \equiv 4\Delta t/S$ with the dimensionless measurement strength $S=64\chi^2\bar{n}\eta_m\Delta t/\kappa$.

The formalism for generating a joint qubit trajectory is quite similar to that of the single qubit case.  We collect a series of measurements $\{r_k\}$ at times $\{t_k\}$, and use these measurements to calculate the conditional density matrices $\{\rho_k^{ij,lm}\}$, where $\{ij, lm\}$ index the computational states.  The diagonal density matrix elements can be calculated using a Bayes' rule, for example:
\begin{equation}
\frac{\rho_k^{00,00} }{\rho_k^{11,11}}= \frac{\rho_0^{00,00} (0)e^{\left[ - \left(r_k +2\right)^2/2\sigma^2\right]}}{\rho_0^{11,11} (0)e^{\left[ - \left(r_k -2\right)^2/2\sigma^2\right]}}.
\label{diags}
\end{equation}

Here, $\sigma$ is the width of the Gaussian histograms, which decreases as $1/\sqrt{\Delta t}$.  The off-diagonal density matrix elements can also be calculated within the same Bayesian formalism.   Neglecting internal losses in the cavity and $T_1$ relaxation, the off-diagonal density matrix terms $\rho_k^{01,10}$ are given by:

\begin{eqnarray}
&& \label{offdiags} \hspace{-0.7cm} |\rho^{ij,lm}_{k}|=  |\rho^{ij,lm}_{0}| \,
\frac{\sqrt{\rho^{ij,ij}_{k}\rho^{lm, lm}_{k}}}  {\sqrt{\rho^{ij,ij}_{0}\rho^{lm,lm}_{0}}} e^{-\gamma_{ij,lm} k\Delta t}  \\
&&\hspace{-0.7cm} \times \text{ exp} \left[-\frac{k\Delta t}{2} \left( 1- \eta_m\right) |V_k^{ij} - V_k^{lm}|^2 \right] \nonumber, 
\end{eqnarray}

\noindent where $V^{i,j}_k$ is the average voltage corresponding to the qubits prepared in the states $i,j$. Here, the first term represents the Bayesian update and includes intrinsic $T_2^*$ dephasing of the matrix element $\gamma_{ij,lm}$; the second term accounts for partial dephasing due to uncollected measurement photons.  Notice that in the case where $V_k^{01} = V_k^{10}$, there is no dephasing of the $\rho^{01,10}$ off-diagonal term due to nonunity $\eta_m$: the odd-parity subspace becomes protected with respect to added noise, and we expect the measurement to probabilistically generate entanglement in the odd-parity subspace.  

Equations (\ref{diags}) and (\ref{offdiags}) provide a mapping $\{r_k\} \mapsto \{\rho_k\}$ at each measurement time $t_k=k\Delta t$. We can also reconstruct the trajectories experimentally using conditional tomography, as described in Section \ref{cond} and elaborated in reference \cite{roch14}. Figure 4C shows such a Baeysian mapping. However, as mentioned earlier, the losses between the cavities need to be accounted for. Thus we used a more refined formalism as explained in reference \cite{roch14} . From this mapping, one can reconstruct the trajectory of a single iteration of the experiment.

 As mentioned before, the main advantage of this cascaded system is its ability to generate entanglement between remote qubits. To quantify this entanglement, we can reconstruct the quantum trajectory of the concurrence, which is a monotone of entanglement. A simplified definition is given by:
 
 \begin{equation}
 C_k=2\max(0,|\rho^{01,10}_k|-\sqrt{\rho_k^{00,00}\rho_k^{11,11}})
 \end{equation}

Concurrence reaches a maximum value of 1 for a maximally-entangled Bell state, and is zero for joint qubit states that cannot be distinguished from separable or from  classically mixed states.  An exemplar quantum trajectory of the joint qubit state, and of the concurrence, are shown in Figure 4D. We see an initial transient during which $C$ is zero, followed by a non-monotonic increase in the $C$ as the joint qubit state stochastically projects towards the entangled manifold, reaching an eventual concurrence of $C\sim0.55$, indicating a highly nonclassical state.

\section{Outlook}

The experiments presented in this review demonstrate precise control and a detailed understanding of the process of continuous quantum measurement of a superconducting qubit.  This knowledge may benefit a wide range of future research directions in quantum control and multi-qubit state estimation \cite{silb05, smit13}.  In this final section we highlight one such research direction: measurement-based quantum feedback.  
 
A continuous measurement record, which contains information about how the qubit state evolves in real time, can be incorporated into a feedback loop for a number of applications including state preparation, state stabilization, and continuous quantum error correction.  Without feedback, a combination of projective measurement and unitary rotation can be used to probabilistically prepare an arbitrary qubit state \cite{john12}.  Using feedback, the measurement result can be used to control a subsequent qubit rotation, allowing for deterministic  state-reset protocols \cite{rist12,camp13} which can be repeated on a timescale much faster than $T_1$.   
 
In the case of weak measurement, it is possible prepare arbitrary states by measurement alone, without applying any subsequent qubit rotations.  Without feedback, state preparation is probabilistic: one simply post-selects an ensemble of trajectories which end up in the desired state.  With feedback, it is possible to prepare an arbitrary qubit state deterministically through adaptive measurement, as recently demonstrated using nitrogen vacancy centers \cite{blok14}. 
 
In addition to state preparation, quantum feedback can also be used to stabilize a qubit state or trajectory.  In reference \cite{vija12}, we used weak measurements and continuous quantum feedback to stabilize Rabi oscillations of a superconducting qubit.  Reference \cite{camp13} demonstrated that stroboscopic projective measurements and feedback can be used to stabilize an arbitrary trajectory, such as Rabi or Ramsey oscillations.  Furthermore, with the with the two-qubit setup from reference \cite{roch14} it should be possible to use feedback to to stabilize an entangled state. 
 
Looking forward, proposals for fault tolerant quantum rely on quantum error correction (QEC) protocols in which a single logical qubit is composed of many physical qubits.  While many QEC schemes such as surface codes rely on discrete projective measurements of syndrome qubits \cite{brav98, chow14, bard14}, a wide body of QEC proposals are based instead on continuous measurement-based quantum feedback \cite{wisebook}.  In these techniques, a single logical qubit is encoded in several physical qubits, and an error syndrome is detected by processing (or `filtering') a continuous measurement signal.  The error signal is used to generate a suitable feedback Hamiltonian which corrects for errors in real time.
 
By tomographically validating individual quantum trajectories, the experiments presented in this review have demonstrated the ability to correctly `filter' a measurement signal for one and two qubit systems. A sensible next step is to build a system of several qubits and attempt to correctly filter an error syndrome.  Then, the following step would be to feed-back on this error syndrome to realize a single logical qubit whose lifetime exceeds that of its constituent physical qubits.  To realize this goal will require a robust multi-qubit architecture and improvements in the measurement quantum efficiency.  Recent experiments \cite{bard14, kell14} have demonstrated that it is possible to individually measure and control $5\text{-}9$ qubits in a planar cQED architecture, and efforts to improve the measurement quantum efficiency are currently underway in a number of different research groups.  Although there are still formidable challenges to overcome, and while the ultimate utility of measurement based QEC in comparison to other methods remains an open question, it seems that an initial demonstration of measurement based QEC may lie on the horizon. 
 
%Traditional protocols for quantum state tomography (QST) of a multi-qubit system are based on a series of fixed rotations, each followed by a strong projective measurement.  The resource requirements for this type of sequential characterization are prohibitive for large numbers of qubits, and result in errors due to state evolution from the strong measurement pulse and any residual qubit couplings which can not be switched off.
% 
%It may be possible to develop tomography protocols based on continuous quantum measurement whose resource requirements scale more favorably with system size than conventional QST protocals.  The group of Ivan Deutsch introduced protocols for state estimation based on repeated continuous weak measurements and continuous driving \cite{silb05,smit13}.  During driven evolution, a single measurement can probe the system in multiple bases, and it is possible to gain an informationally complete measurement record from a single ensemble of identically prepared qubits.  We are currently developing similar techniques which involve simultaneous continuous measurements of multiple superconducting qubits.

\section*{Acknowledgements}

This research was supported in part by the Army Research Office, Office of
Naval Research and the Office of the Director of National Intelligence (ODNI), Intelligence Advanced Research Projects Activity (IARPA), through the Army Research Office. All statements of fact, opinion or conclusions contained herein are those of the authors and should not be construed as representing the official views or policies of IARPA, the ODNI or the US government. MES acknowledges support from the Fannie and John Hertz Foundation.

\label{out}

  \bibliographystyle{elsarticle-num} 
  \bibliography{trajreview,pqs_references}

\begin{thebibliography}{10}
\expandafter\ifx\csname url\endcsname\relax
  \def\url#1{\texttt{#1}}\fi
\expandafter\ifx\csname urlprefix\endcsname\relax\def\urlprefix{URL }\fi
\expandafter\ifx\csname href\endcsname\relax
  \def\href#1#2{#2} \def\path#1{#1}\fi

\bibitem{carmbook}
H.~J. Carmichael, An Open Systems Approach to Quantum Optics, Springer, Berlin,
  1993.

\bibitem{gardinerbook}
C.~Gardiner, P.~Zoller, Quantum Noise, Springer, 2004.

\bibitem{dali92}
J.~Dalibard, Y.~Castin, K.~M{\o}lmer, {Wave-function approach to dissipative
  processes in quantum optics} 68 (1992) 580--583.
\newblock \href {http://dx.doi.org/10.1103/PhysRevLett.68.580}
  {\path{doi:10.1103/PhysRevLett.68.580}}.

\bibitem{gard92}
C.~Gardiner, A.~Parkins, P.~Zoller, {Wave-function quantum stochastic
  differential equations and quantum-jump simulation methods} 46 (1992)
  4363--4381.
\newblock \href {http://dx.doi.org/10.1103/PhysRevA.46.4363}
  {\path{doi:10.1103/PhysRevA.46.4363}}.

\bibitem{scha95}
R.~Schack, T.~A. Brun, I.~C. Percival, Quantum state diffusion, localization
  and computation, J. Phys. A. 28 (1995) 5401--5413.

\bibitem{wisebook}
H.~M. Wiseman, G.~J. Milburn, Quantum Measurement and Control, Cambridge
  University Press, 2010.

\bibitem{brun02}
T.~A. Brun, {A simple model of quantum trajectories}, Am. J. Phys. 70 (2002) 7.

\bibitem{gisi84}
N.~Gisin, {Quantum Measurements and Stochastic Processes}, Physical Review
  Letters 52~(19) (1984) 1657--1660.
\newblock \href {http://dx.doi.org/10.1103/PhysRevLett.52.1657}
  {\path{doi:10.1103/PhysRevLett.52.1657}}.

\bibitem{dios88}
L.~Diosi, {Quantum stochastic processes as models for state vector reduction},
  Journal of Physics A: Mathematical and General 21~(13) (1988) 2885--2898.
\newblock \href {http://dx.doi.org/10.1088/0305-4470/21/13/013}
  {\path{doi:10.1088/0305-4470/21/13/013}}.

\bibitem{gisi92}
N.~Gisin, I.~C. Percival, {The quantum-state diffusion model applied to open
  systems}, Journal of Physics A: Mathematical and General 25~(21) (1992)
  5677--5691.
\newblock \href {http://dx.doi.org/10.1088/0305-4470/25/21/023}
  {\path{doi:10.1088/0305-4470/25/21/023}}.

\bibitem{grif84}
R.~B. Griffiths, {Consistent histories and the interpretation of quantum
  mechanics}, Journal of Statistical Physics 36~(1-2) (1984) 219--272.
\newblock \href {http://dx.doi.org/10.1007/BF01015734}
  {\path{doi:10.1007/BF01015734}}.

\bibitem{nago86}
W.~Nagourney, J.~Sandberg, H.~Dehmelt, {Shelved optical electron amplifier:
  Observation of quantum jumps}, Physical Review Letters 56~(26) (1986)
  2797--2799.
\newblock \href {http://dx.doi.org/10.1103/PhysRevLett.56.2797}
  {\path{doi:10.1103/PhysRevLett.56.2797}}.

\bibitem{saut86}
T.~Sauter, W.~Neuhauser, R.~Blatt, P.~Toschek, {Observation of Quantum Jumps},
  Physical Review Letters 57~(14) (1986) 1696--1698.
\newblock \href {http://dx.doi.org/10.1103/PhysRevLett.57.1696}
  {\path{doi:10.1103/PhysRevLett.57.1696}}.

\bibitem{berg86}
J.~Bergquist, R.~Hulet, W.~Itano, D.~Wineland, {Observation of Quantum Jumps in
  a Single Atom}, Physical Review Letters 57~(14) (1986) 1699--1702.
\newblock \href {http://dx.doi.org/10.1103/PhysRevLett.57.1699}
  {\path{doi:10.1103/PhysRevLett.57.1699}}.

\bibitem{vija11}
R.~Vijay, D.~H. Slichter, I.~Siddiqi, {Observation of Quantum Jumps in a
  Superconducting Artificial Atom}, Physical Review Letters 106~(11) (2011)
  110502.
\newblock \href {http://dx.doi.org/10.1103/PhysRevLett.106.110502}
  {\path{doi:10.1103/PhysRevLett.106.110502}}.

\bibitem{guer07}
C.~Guerlin, J.~Bernu, S.~Del\'{e}glise, C.~Sayrin, S.~Gleyzes, S.~Kuhr,
  M.~Brune, J.-M. Raimond, S.~Haroche, {Progressive field-state collapse and
  quantum non-demolition photon counting}, Nature 448~(7156) (2007) 889--93.
\newblock \href {http://dx.doi.org/10.1038/nature06057}
  {\path{doi:10.1038/nature06057}}.

\bibitem{hood00}
C.~J. Hood, {The Atom-Cavity Microscope: Single Atoms Bound in Orbit by Single
  Photons}, Science 287~(5457) (2000) 1447--1453.
\newblock \href {http://dx.doi.org/10.1126/science.287.5457.1447}
  {\path{doi:10.1126/science.287.5457.1447}}.

\bibitem{koro99}
A.~Korotkov, {Continuous quantum measurement of a double dot}, Physical Review
  B 60~(8) (1999) 5737--5742.
\newblock \href {http://dx.doi.org/10.1103/PhysRevB.60.5737}
  {\path{doi:10.1103/PhysRevB.60.5737}}.

\bibitem{goan01}
H.-S. Goan, G.~Milburn, H.~Wiseman, H.~{Bi Sun}, {Continuous quantum
  measurement of two coupled quantum dots using a point contact: A quantum
  trajectory approach}, Physical Review B 63~(12) (2001) 125326.
\newblock \href {http://dx.doi.org/10.1103/PhysRevB.63.125326}
  {\path{doi:10.1103/PhysRevB.63.125326}}.

\bibitem{sukh07}
E.~V. Sukhorukov, A.~N. Jordan, S.~Gustavsson, R.~Leturcq, T.~Ihn, K.~Ensslin,
  {Conditional statistics of electron transport in interacting nanoscale
  conductors}, Nature Physics 3~(4) (2007) 243--247.
\newblock \href {http://dx.doi.org/10.1038/nphys564}
  {\path{doi:10.1038/nphys564}}.

\bibitem{gamb08}
J.~Gambetta, A.~Blais, M.~Boissonneault, A.~A. Houck, D.~I. Schuster, S.~M.
  Girvin, {Quantum trajectory approach to circuit QED: Quantum jumps and the
  Zeno effect}, Physical Review A - Atomic, Molecular, and Optical Physics
  77~(1) (2008) 012112.
\newblock \href {http://dx.doi.org/10.1103/PhysRevA.77.012112}
  {\path{doi:10.1103/PhysRevA.77.012112}}.

\bibitem{koro11}
A.~N. Korotkov, {Quantum Bayesian approach to circuit QED measurement} (2011)
  arXiv:1111.4016.

\bibitem{hatr13}
M.~Hatridge, S.~Shankar, M.~Mirrahimi, F.~Schackert, K.~Geerlings, T.~Brecht,
  K.~M. Sliwa, B.~Abdo, L.~Frunzio, S.~M. Girvin, R.~J. Schoelkopf, M.~H.
  Devoret, {Quantum back-action of an individual variable-strength
  measurement}, Science 339 (2013) 178--81.
\newblock \href {http://dx.doi.org/10.1126/science.1226897}
  {\path{doi:10.1126/science.1226897}}.

\bibitem{camp14}
P.~Campagne-Ibarcq, L.~Bretheau, E.~Flurin, A.~Auff\`{e}ves, F.~Mallet,
  B.~Huard, {Observing Interferences between Past and Future Quantum States in
  Resonance Fluorescence}, Physical Review Letters 112~(18) (2014) 180402.
\newblock \href {http://dx.doi.org/10.1103/PhysRevLett.112.180402}
  {\path{doi:10.1103/PhysRevLett.112.180402}}.

\bibitem{murc13}
K.~W. Murch, S.~J. Weber, C.~Macklin, I.~Siddiqi, {Observing single quantum
  trajectories of a superconducting quantum bit}, Nature 502~(7470) (2013)
  211--4.
\newblock \href {http://dx.doi.org/10.1038/nature12539}
  {\path{doi:10.1038/nature12539}}.

\bibitem{webe14}
S.~J. Weber, A.~Chantasri, J.~Dressel, A.~N. Jordan, K.~W. Murch, I.~Siddiqi,
  {Mapping the optimal route between two quantum states}, Nature 511 (2014)
  570--573.
\newblock \href {http://dx.doi.org/10.1038/nature13559}
  {\path{doi:10.1038/nature13559}}.

\bibitem{roch14}
N.~Roch, M.~Schwartz, F.~Motzoi, C.~Macklin, R.~Vijay, A.~Eddins, A.~Korotkov,
  K.~Whaley, M.~Sarovar, I.~Siddiqi, {Observation of Measurement-Induced
  Entanglement and Quantum Trajectories of Remote Superconducting Qubits},
  Physical Review Letters 112~(17) (2014) 170501.
\newblock \href {http://dx.doi.org/10.1103/PhysRevLett.112.170501}
  {\path{doi:10.1103/PhysRevLett.112.170501}}.

\bibitem{tan14}
D.~Tan, S.~Weber, I.~Siddiqi, K.~M{\o}lmer, K.~W. Murch, {Prediction and
  retrodiction for a continuously monitored superconducting qubit} (2014)
  arXiv:1409.0510.

\bibitem{sayr11}
C.~Sayrin, I.~Dotsenko, X.~Zhou, B.~Peaudecerf, T.~Rybarczyk, S.~Gleyzes,
  P.~Rouchon, M.~Mirrahimi, H.~Amini, M.~Brune, J.-M. Raimond, S.~Haroche,
  {Real-time quantum feedback prepares and stabilizes photon number states},
  Nature 477~(7362) (2011) 73--7.
\newblock \href {http://dx.doi.org/10.1038/nature10376}
  {\path{doi:10.1038/nature10376}}.

\bibitem{vija12}
R.~Vijay, C.~Macklin, D.~H. Slichter, S.~J. Weber, K.~W. Murch, R.~Naik, A.~N.
  Korotkov, I.~Siddiqi, {Stabilizing Rabi oscillations in a superconducting
  qubit using quantum feedback}, Nature 490~(7418) (2012) 77--80.
\newblock \href {http://dx.doi.org/10.1038/nature11505}
  {\path{doi:10.1038/nature11505}}.

\bibitem{dela14}
G.~de~Lange, D.~Rist\`{e}, M.~Tiggelman, C.~Eichler, L.~Tornberg, G.~Johansson,
  A.~Wallraff, R.~Schouten, L.~DiCarlo, {Reversing Quantum Trajectories with
  Analog Feedback}, Physical Review Letters 112~(8) (2014) 080501.
\newblock \href {http://dx.doi.org/10.1103/PhysRevLett.112.080501}
  {\path{doi:10.1103/PhysRevLett.112.080501}}.

\bibitem{groe13}
J.~P. Groen, D.~Rist\`{e}, L.~Tornberg, J.~Cramer, P.~C. {De Groot}, T.~Picot,
  G.~Johansson, L.~Dicarlo, {Partial-measurement backaction and nonclassical
  weak values in a superconducting circuit}, Physical Review Letters 111.
\newblock \href {http://dx.doi.org/10.1103/PhysRevLett.111.090506}
  {\path{doi:10.1103/PhysRevLett.111.090506}}.

\bibitem{blok14}
M.~S. Blok, C.~Bonato, M.~L. Markham, D.~J. Twitchen, V.~V. Dobrovitski,
  R.~Hanson, {Manipulating a qubit through the backaction of sequential partial
  measurements and real-time feedback}, Nature Physics 10~(3) (2014) 189--193.
\newblock \href {http://dx.doi.org/10.1038/nphys2881}
  {\path{doi:10.1038/nphys2881}}.

\bibitem{koch07}
J.~Koch, T.~Yu, J.~Gambetta, A.~Houck, D.~Schuster, J.~Majer, A.~Blais,
  M.~Devoret, S.~Girvin, R.~Schoelkopf, {Charge-insensitive qubit design
  derived from the Cooper pair box}, Physical Review A 76~(4) (2007) 042319.
\newblock \href {http://dx.doi.org/10.1103/PhysRevA.76.042319}
  {\path{doi:10.1103/PhysRevA.76.042319}}.

\bibitem{megr12}
A.~Megrant, C.~Neill, R.~Barends, B.~Chiaro, Y.~Chen, L.~Feigl, J.~Kelly,
  E.~Lucero, M.~Mariantoni, P.~J.~J. O'Malley, D.~Sank, A.~Vainsencher,
  J.~Wenner, T.~C. White, Y.~Yin, J.~Zhao, C.~J. Palmstro̸m, J.~M. Martinis,
  A.~N. Cleland, {Planar superconducting resonators with internal quality
  factors above one million}, Applied Physics Letters 100~(11) (2012) 113510.
\newblock \href {http://dx.doi.org/10.1063/1.3693409}
  {\path{doi:10.1063/1.3693409}}.

\bibitem{chan13}
J.~B. Chang, M.~R. Vissers, A.~D. C{\'o}rcoles, M.~Sandberg, J.~Gao, D.~W.
  Abraham, J.~M. Chow, J.~M. Gambetta, M.~{Beth Rothwell}, G.~A. Keefe,
  M.~Steffen, D.~P. Pappas, {Improved superconducting qubit coherence using
  titanium nitride}, Applied Physics Letters 103~(1) (2013) 012602.
\newblock \href {http://dx.doi.org/10.1063/1.4813269}
  {\path{doi:10.1063/1.4813269}}.

\bibitem{paik11}
H.~Paik, D.~I. Schuster, L.~S. Bishop, G.~Kirchmair, G.~Catelani, A.~P. Sears,
  B.~R. Johnson, M.~J. Reagor, L.~Frunzio, L.~I. Glazman, S.~M. Girvin, M.~H.
  Devoret, R.~J. Schoelkopf, {Observation of High Coherence in Josephson
  Junction Qubits Measured in a Three-Dimensional Circuit QED Architecture},
  Physical Review Letters 107~(24) (2011) 240501.
\newblock \href {http://dx.doi.org/10.1103/PhysRevLett.107.240501}
  {\path{doi:10.1103/PhysRevLett.107.240501}}.

\bibitem{blai04}
A.~Blais, R.-S. Huang, A.~Wallraff, S.~Girvin, R.~Schoelkopf, {Cavity quantum
  electrodynamics for superconducting electrical circuits: An architecture for
  quantum computation}, Physical Review A 69~(6) (2004) 062320.
\newblock \href {http://dx.doi.org/10.1103/PhysRevA.69.062320}
  {\path{doi:10.1103/PhysRevA.69.062320}}.

\bibitem{bragbook}
V.~B. Braginsky, F.~Y. Khalili, {Qunatum Measurement}, Cambridge University
  Press, 1992.

\bibitem{berg10}
N.~Bergeal, F.~Schackert, M.~Metcalfe, R.~Vijay, V.~E. Manucharyan, L.~Frunzio,
  D.~E. Prober, R.~J. Schoelkopf, S.~M. Girvin, M.~H. Devoret,
  {Phase-preserving amplification near the quantum limit with a Josephson ring
  modulator}, Nature 465~(7294) (2010) 64--8.
\newblock \href {http://dx.doi.org/10.1038/nature09035}
  {\path{doi:10.1038/nature09035}}.

\bibitem{cave82}
C.~Caves, {Quantum limits on noise in linear amplifiers}, Physical Review D
  26~(8) (1982) 1817--1839.
\newblock \href {http://dx.doi.org/10.1103/PhysRevD.26.1817}
  {\path{doi:10.1103/PhysRevD.26.1817}}.

\bibitem{hatr11}
M.~Hatridge, R.~Vijay, D.~H. Slichter, J.~Clarke, I.~Siddiqi, {Dispersive
  magnetometry with a quantum limited SQUID parametric amplifier}, Physical
  Review B 83~(13) (2011) 134501.
\newblock \href {http://dx.doi.org/10.1103/PhysRevB.83.134501}
  {\path{doi:10.1103/PhysRevB.83.134501}}.

\bibitem{jaco06}
K.~Jacobs, D.~Steck, A straightforward introduction to continuous quantum
  measurement, Contemporary Physics 47 (2006) 279.

\bibitem{chan13b}
A.~Chantasri, J.~Dressel, A.~N. Jordan, {Action principle for continuous
  quantum measurement}, Physical Review A 88~(4) (2013) 042110.
\newblock \href {http://dx.doi.org/10.1103/PhysRevA.88.042110}
  {\path{doi:10.1103/PhysRevA.88.042110}}.

\bibitem{wise02}
H.~M. Wiseman, Weak values, quantum trajectories, and the cavity-qed experiment
  on wave-particle correlation, Phys. Rev. A 65 (2002) 032111.
\newblock \href {http://dx.doi.org/10.1103/PhysRevA.65.032111}
  {\path{doi:10.1103/PhysRevA.65.032111}}.

\bibitem{gamm13}
S.~Gammelmark, B.~Julsgaard, K.~M\o{}lmer, Past quantum states of a monitored
  system, Phys. Rev. Lett. 111 (2013) 160401.
\newblock \href {http://dx.doi.org/10.1103/PhysRevLett.111.160401}
  {\path{doi:10.1103/PhysRevLett.111.160401}}.

\bibitem{mank09}
M.~Tsang, Time-symmetric quantum theory of smoothing, Phys. Rev. Lett. 102
  (2009) 250403.
\newblock \href {http://dx.doi.org/10.1103/PhysRevLett.102.250403}
  {\path{doi:10.1103/PhysRevLett.102.250403}}.

\bibitem{mank09pra}
M.~Tsang, Optimal waveform estimation for classical and quantum systems via
  time-symmetric smoothing, Phys. Rev. A 80 (2009) 033840.
\newblock \href {http://dx.doi.org/10.1103/PhysRevA.80.033840}
  {\path{doi:10.1103/PhysRevA.80.033840}}.

\bibitem{mank11}
M.~Tsang, H.~M. Wiseman, C.~M. Caves, Fundamental quantum limit to waveform
  estimation, Phys. Rev. Lett. 106 (2011) 090401.
\newblock \href {http://dx.doi.org/10.1103/PhysRevLett.106.090401}
  {\path{doi:10.1103/PhysRevLett.106.090401}}.

\bibitem{arme09}
M.~A. Armen, A.~E. Miller, H.~Mabuchi, Spontaneous dressed-state polarization
  in the strong driving regime of cavity qed, Phys. Rev. Lett. 103 (2009)
  173601.
\newblock \href {http://dx.doi.org/10.1103/PhysRevLett.103.173601}
  {\path{doi:10.1103/PhysRevLett.103.173601}}.

\bibitem{whea10}
T.~A. Wheatley, D.~W. Berry, H.~Yonezawa, D.~Nakane, H.~Arao, D.~T. Pope, T.~C.
  Ralph, H.~M. Wiseman, A.~Furusawa, E.~H. Huntington, Adaptive optical phase
  estimation using time-symmetric quantum smoothing, Phys. Rev. Lett. 104
  (2010) 093601.
\newblock \href {http://dx.doi.org/10.1103/PhysRevLett.104.093601}
  {\path{doi:10.1103/PhysRevLett.104.093601}}.

\bibitem{ryba14}
T.~Rybarczyk, S.~Gerlich, B.~Peaudecerf, M.~Penasa, B.~Julsgaard, K.~M{\o}lmer,
  S.~Gleyzes, M.~Brune, J.~M. Raimond, S.~Haroche, I.~Dotsenko, Past quantum
  state analysis of the photon number evolution in a cavity (2014)
  arXiv:1409.0958.

\bibitem{will08b}
N.~Williams, A.~Jordan, {Entanglement genesis under continuous parity
  measurement}, Physical Review A 78~(6) (2008) 062322.

\bibitem{silb05}
A.~Silberfarb, P.~Jessen, I.~Deutsch, {Quantum State Reconstruction via
  Continuous Measurement}, Physical Review Letters 95~(3) (2005) 030402.
\newblock \href {http://dx.doi.org/10.1103/PhysRevLett.95.030402}
  {\path{doi:10.1103/PhysRevLett.95.030402}}.

\bibitem{smit13}
A.~Smith, C.~Riofr\'{\i}o, B.~Anderson, H.~Sosa-Martinez, I.~Deutsch,
  P.~Jessen, {Quantum state tomography by continuous measurement and compressed
  sensing}, Physical Review A 87~(3) (2013) 030102.
\newblock \href {http://dx.doi.org/10.1103/PhysRevA.87.030102}
  {\path{doi:10.1103/PhysRevA.87.030102}}.

\bibitem{john12}
J.~E. Johnson, C.~Macklin, D.~H. Slichter, R.~Vijay, E.~B. Weingarten,
  J.~Clarke, I.~Siddiqi, {Heralded State Preparation in a Superconducting
  Qubit}, Physical Review Letters 109~(5) (2012) 050506.
\newblock \href {http://dx.doi.org/10.1103/PhysRevLett.109.050506}
  {\path{doi:10.1103/PhysRevLett.109.050506}}.

\bibitem{rist12}
D.~Rist\`{e}, C.~C. Bultink, K.~W. Lehnert, L.~DiCarlo, {Feedback Control of a
  Solid-State Qubit Using High-Fidelity Projective Measurement}, Physical
  Review Letters 109~(24) (2012) 240502.
\newblock \href {http://dx.doi.org/10.1103/PhysRevLett.109.240502}
  {\path{doi:10.1103/PhysRevLett.109.240502}}.

\bibitem{camp13}
P.~Campagne-Ibarcq, E.~Flurin, N.~Roch, D.~Darson, P.~Morfin, M.~Mirrahimi,
  M.~H. Devoret, F.~Mallet, B.~Huard, {Persistent Control of a Superconducting
  Qubit by Stroboscopic Measurement Feedback}, Physical Review X 3~(2) (2013)
  021008.
\newblock \href {http://dx.doi.org/10.1103/PhysRevX.3.021008}
  {\path{doi:10.1103/PhysRevX.3.021008}}.

\bibitem{brav98}
S.~Bravyi, A.~Kitaev, Quantum codes on a lattice with boundary (1998)
  arXiv:quant--ph/9811052.

\bibitem{chow14}
J.~M. Chow, J.~M. Gambetta, E.~Magesan, D.~W. Abraham, A.~W. Cross, B.~R.
  Johnson, N.~A. Masluk, C.~A. Ryan, J.~A. Smolin, S.~J. Srinivasan,
  M.~Steffen, {Implementing a strand of a scalable fault-tolerant quantum
  computing fabric}, Nature communications 5 (2014) 4015.
\newblock \href {http://dx.doi.org/10.1038/ncomms5015}
  {\path{doi:10.1038/ncomms5015}}.

\bibitem{bard14}
R.~Barends, J.~Kelly, A.~Megrant, A.~Veitia, D.~Sank, E.~Jeffrey, T.~C. White,
  J.~Mutus, A.~G. Fowler, B.~Campbell, Y.~Chen, Z.~Chen, B.~Chiaro,
  A.~Dunsworth, C.~Neill, P.~O'Malley, P.~Roushan, A.~Vainsencher, J.~Wenner,
  A.~N. Korotkov, A.~N. Cleland, J.~M. Martinis, {Superconducting quantum
  circuits at the surface code threshold for fault tolerance}, Nature
  508~(7497) (2014) 500--3.
\newblock \href {http://dx.doi.org/10.1038/nature13171}
  {\path{doi:10.1038/nature13171}}.

\bibitem{kell14}
J.~Kelly, R.~Barends, A.~G. Fowler, A.~Megrant, E.~Jeffrey, T.~C. White,
  D.~Sank, J.~Y. Mutus, B.~Campbell, Y.~Chen, Z.~Chen, B.~Chiaro, A.~Dunsworth,
  I.~C. Hoi, C.~Neill, P.~J.~J. O'Malley, C.~Quintana, P.~Roushan,
  A.~Vainsencher, J.~Wenner, A.~N. Cleland, J.~M. Martinis, {State preservation
  by repetitive error detection in a superconducting quantum circuit} (2014)
  arXiv:1411.7403.

\end{thebibliography}




%%\bibitem{}

%%\end{thebibliography}
\end{document}